\documentclass[seceq]{ptptex}
\usepackage{wrapft}
\usepackage{bm}

\title{
Misleading Coupling Unification
and Lifshitz Type Gauge Theory}

\author{
Yoshiharu \textsc{Kawamura}\footnote{E-mail: haru@azusa.shinshu-u.ac.jp}
}

\inst{
Department of Physics, Shinshu University, Matsumoto 390-8621, Japan
}

\recdate{
June 20, 2009}

\abst{
We propose a new grand unification scenario for ensuring proton stability. 
Our scenario is based on the idea that the proton decay problem 
is an artificial one, which is caused from the identification of the gauge coupling unification scale 
with the grand unification scale or the grand unified symmetry breaking scale.
We discuss a Lifshitz type gauge theory as a candidate
to realize our scenario.
}

\begin{document}

\maketitle

\section{Introduction}

Grand unification is an attractive idea and enables the unification of forces 
and the (partial) unification of quarks and leptons in each family.\cite{GUT}
The introduction of supersymmetry (SUSY) increases the reality of grand unification.
In the minimal supersymmetric standard model (MSSM),
the gauge couplings meet at $M_X = 2.1 \times 10^{16}$GeV
if the superpartners and Higgs particles exist below or around $O(1)$TeV.\cite{unif}$^,$\footnote{
The gauge coupling unification based on SUSY was predicted in Refs.~\citen{unif0}.}
It is natural to guess that the grand unification occurs at $M_X$ and physics above $M_X$ can be described 
as a supersymmetric grand unified theory (SUSY GUT).
This scenario is very attractive, but, in general, it suffers from problems related to Higgs multiplets.
The typical one is the proton decay problem\cite{Proton}
in the minimal SUSY GUT.\cite{SUSYGUT}

The proton decay problem comes from a significant contribution from the dimension 5 operators.
Stronger constraints on the colored Higgs mass $M_{H_C}$ and the sfermion mass $m_{\tilde{f}}$
have been obtained (e.g.~$M_{H_C} > 6.5 \times 10^{16}$GeV for $m_{\tilde{f}} < 1$TeV)
from analysis including a Higgsino dressing diagram with right-handed matter fields in the minimal SUSY $SU(5)$ GUT.\cite{dim5}
Extra particles such as $X$, $Y$ gauge bosons, the colored Higgs bosons and their superpartners
are expected to acquire heavy masses of $O(M_U)$ after the breakdown of grand unified symmetry.
Here $M_U$ is the grand unification scale or the grand unified symmetry breaking scale.
The constraints contradict with the relation $M_{H_C} = O(M_U)$ if $M_U$ is identified with 
the gauge coupling unification scale $M_X$.
{}From this observation, we come across the conjecture that
{\it the proton stability can be realized to be compatible with the grand unification,
if $M_X$ is not directly related to $M_U$ and the order of $M_U$ is bigger than that of $M_X$.}
It might lead to the idea that gauge couplings agree with accidentally at $M_X$.

In this paper, we propose a new grand unification scenario for ensuring proton stability 
on the basis of the above conjecture and the standpoint that {\it the scale $M_X$ has a physical meaning
beyond the fact that the agreement of gauge couplings occurs accidentally there.}
We discuss a Lifshitz type gauge theory as a candidate to realize our scenario.

This paper is organized as follows.
In the next section, we elaborate our scenario.
In \S 3, we discuss a candidate to realize our scenario.
In \S 4, we present conclusions and discussion.
In Appendix A, we study radiative corrections using a Lifshitz type abelian gauge theory.

\section{Our scenario}

The proton stability can be threatened through contributions from 
the dimension 5 operators including the colored Higgsinos 
in the minimal SUSY $SU(5)$ GUT.
Stronger constraints on the colored Higgs mass $M_{H_C}$ and the sfermion mass $m_{\tilde{f}}$
have been obtained (e.g.~$M_{H_C} > 6.5 \times 10^{16}$GeV for $m_{\tilde{f}} < 1$TeV)
from analysis including a Higgsino dressing diagram with right-handed matter fields.\cite{dim5}$^,$\footnote{
In Refs.~\citen{BP&S}, an intriguing possibility is pointed out that there are some uncertainties 
in estimating the dimension 5 operators and even the minimal SUSY $SU(5)$ GUT survives.
}
Extra particles such as $X$, $Y$ gauge bosons, the colored Higgs bosons and their superpartners
usually acquire heavy masses of $O(M_U)$ after the breakdown of grand unified symmetry at $M_U$.
The root of proton decay problem stems from the relation $M_{H_C} = O(M_U)$
with the identification between the grand unified symmetry breaking scale $M_U$ 
and the gauge coupling unification scale $M_X = 2.1 \times 10^{16}$GeV.
Conventionally, one tackles the problem through the extension of the model 
leaving the identification untouched.

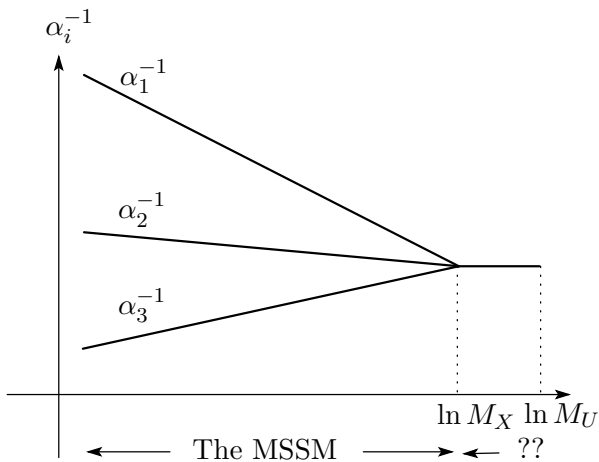
\begin{wrapfigure}{l}{7.5cm}
\label{F1.alphai}
\begin{center}
\unitlength 0.1in
\begin{picture}( 29.4000, 23.4000)(  2.2000,-23.7000)
%
\special{pn 8}%
\special{pa 3012 1368}%
\special{pa 3012 2028}%
\special{dt 0.045}%
%
\special{pn 13}%
\special{pa 622 366}%
\special{pa 2584 1368}%
\special{fp}%
%
\special{pn 13}%
\special{pa 622 1190}%
\special{pa 2584 1368}%
\special{fp}%
%
\special{pn 13}%
\special{pa 616 1800}%
\special{pa 2584 1368}%
\special{fp}%
%
\special{pn 13}%
\special{pa 2576 1368}%
\special{pa 3006 1368}%
\special{fp}%
\put(4.2000,-2.0000){\makebox(0,0)[lb]{$\alpha^{-1}_i$}}%
\put(24.7000,-22.1800){\makebox(0,0)[lb]{$\ln M_X$}}%
\put(29.2700,-22.1100){\makebox(0,0)[lb]{$\ln M_U$}}%
%
\special{pn 8}%
\special{pa 2084 2340}%
\special{pa 2548 2340}%
\special{fp}%
\special{sh 1}%
\special{pa 2548 2340}%
\special{pa 2482 2320}%
\special{pa 2496 2340}%
\special{pa 2482 2360}%
\special{pa 2548 2340}%
\special{fp}%
%
\special{pn 8}%
\special{pa 1072 2340}%
\special{pa 644 2340}%
\special{fp}%
\special{sh 1}%
\special{pa 644 2340}%
\special{pa 710 2360}%
\special{pa 696 2340}%
\special{pa 710 2320}%
\special{pa 644 2340}%
\special{fp}%
\put(11.9000,-23.8000){\makebox(0,0)[lb]{The MSSM}}%
%
\special{pn 8}%
\special{pa 2794 2346}%
\special{pa 2612 2350}%
\special{fp}%
\special{sh 1}%
\special{pa 2612 2350}%
\special{pa 2678 2368}%
\special{pa 2664 2348}%
\special{pa 2678 2328}%
\special{pa 2612 2350}%
\special{fp}%
\put(28.9000,-23.7000){\makebox(0,0)[lb]{??}}%
%
\special{pn 8}%
\special{pa 2576 1368}%
\special{pa 2576 2028}%
\special{dt 0.045}%
%
\special{pn 8}%
\special{pa 490 2370}%
\special{pa 490 270}%
\special{fp}%
\special{sh 1}%
\special{pa 490 270}%
\special{pa 470 338}%
\special{pa 490 324}%
\special{pa 510 338}%
\special{pa 490 270}%
\special{fp}%
%
\special{pn 8}%
\special{pa 220 2040}%
\special{pa 3160 2040}%
\special{fp}%
\special{sh 1}%
\special{pa 3160 2040}%
\special{pa 3094 2020}%
\special{pa 3108 2040}%
\special{pa 3094 2060}%
\special{pa 3160 2040}%
\special{fp}%
\put(8.0000,-11.6000){\makebox(0,0)[lb]{$\alpha^{-1}_2$}}%
\put(8.0000,-16.6000){\makebox(0,0)[lb]{$\alpha^{-1}_3$}}%
\put(8.1000,-4.6000){\makebox(0,0)[lb]{$\alpha^{-1}_1$}}%
\end{picture}%
\end{center}
\caption{Running of $\alpha^{-1}_i$.}
\end{wrapfigure}
We explore a new possibility of constructing a problem-free model.
The underlying idea of our scenario is that {\it the gauge coupling unification at $M_X$ 
provides misleading information that the gauge symmetries are also unified at $M_X$.
The proton stability can be realized to be compatible with the grand unification,
if the grand unified symmetry breaking scale is not $M_X$ but a bigger than that.
The scale $M_X$ has a physical meaning
beyond the fact that the agreement of gauge couplings occurs accidentally there.}
The outline of our scenario is as follows.
A GUT describes physics above $M_U$, 
which is bigger than $M_X$ to suppress the prodon decay processes sufficiently. 
After the breakdown of grand unified symmetry, 
the running of standard model (SM) gauge couplings $g_i$, in general, differs.
In this case, it is necessary to explain why $g_i$ meet again at $M_X$.
It might occur due to a decoupling of extra particles.
In this paper, we explore a more exotic scenario that $g_i$ run (or stand) together
from $M_U$ to $M_X$ resulted from a specific dynamics.\footnote{
The idea that the proton decay rate is suppressed by the increase of the unification scale
has been proposed with interesting examples realized by the existence of extra 
vector-like split multiples.\cite{magic}
In particular, the pattern of running of gauge couplings in $\lq\lq$fake unification" is same as that in our scenario,
but the causes are different, i.e., the runnings are deeply related to 
the existence of extra particles in theirs and a specific dynamics in ours.
}
Then the stucture of theory changes around $M_X$ as a result of a phase transition 
irrelevant to the grand unification
and the different running of $g_i$ starts there.
A typical running of gauge couplings is depicted in Fig.~1.
The $\alpha_{i}$ $(i=1,2,3)$ are structure constants constructed from $g_i$.
Through two kinds of phase transitions, the MSSM will be derived in the following,
\begin{eqnarray}
\int dt d^D{\bm{x}} {\mathcal{L}}_{U} \stackrel{M_U}{\longrightarrow} \int dt d^{\tilde{D}}{\bm{x}} {\mathcal{L}}_{I}
\stackrel{M_X}{\longrightarrow} \int d^4x {\mathcal{L}}_{\tiny{\rm MSSM}}~,
\label{F1}
\end{eqnarray}
where ${\mathcal{L}}_{U}$, $\mathcal{L}_{I}$ and ${\mathcal{L}}_{\tiny{\rm MSSM}}$ are 
the Lagrangian density of GUT, an intermediate theory and the MSSM
and we have left the possibility that the structure of space-time is also changed.

For our scenario, it is natural to ask the following questions:\\
1. What kind of GUT describes physics above $M_U$?\\
2. How is the grand unified symmetry broken down to a smaller one at $M_U$?\\
3. What kind of theory describes physics at the region between $M_U$ and $M_X$?\\
4. What kind of dynamics makes $g_i$ to run (or stand) together from $M_U$ to $M_X$?\\
5. What kind of phase transition occurs to derive the MSSM around $M_X$?\\
Unfortunately, we presently have no definite answers to these questions, although
we come some conjectures to mind.
Let us look for possible answers by taking the fourth question as a clue.
The simplest possibility is that any gauge couplings do not (almost) run between $M_U$ and $M_X$, i.e.,
the beta functions of $g_i$ (almost) vanish.
We expect that it can stem from a specific superrenormalizable interactions.
Then a phase transition occurs and a theory with superrenormalizable interactions changes
into that with renormalizable interactions.
We take up challenges for this possibility inspired by models with a Lifshitz type fixed point.\cite{Horava,LFT}
For the first and the third questions, 
we expect that the theory has a Lifshitz type fixed point with $z > 1$ above $M_X$,
it has a grand unified symmetry above $M_U$ and the SM symmetry below $M_U$,
and it changes into a renormalizable theory with $z = 1$ like the MSSM around $M_X$.
In this case, the Poincare invariance and SUSY are regarded as accidental symmetries
which emerge in the lower energy scale less than $M_X$ or in the infrared (IR) region.\footnote{
In Refs.~\citen{Anselmi1}, properties and renormalizability for quantum field theories with Lorentz symmetry breaking terms 
have been studied intensively under $\lq\lq$weighted power counting".
Furthermore the extensions of the SM have been proposed in this framework.\cite{Anselmi2}
}

\section{A candidate}

We expect that a specific gauge theory possesses a fixed point with anisotropic scaling characterized 
by dynamical critical exponent $z > 1$ and realizes our scenario, 
but we need a more careful consideration to construct a complete theory.
Here, we present a candidate and discuss its features.

\subsection{Preparations}

Space-time is assumed to be factorized into a product of $D$-dimensional Euclidean space $\bm{R}^D$ and a time $\bm{R}$,
whose coordinates are denoted by $x^{i}$ $(i = 1, \cdots, D)$ (or $\bm{x}$) and $t$, respectively.
In this paper, we study mostly the case with $D=3$.
The dimensions of $x^{i}$ and $t$ are defined as
\begin{eqnarray}
[x^i] = -1~, ~~ [t]=-z~,
\label{dim1}
\end{eqnarray}
where $z$ is the dynamical critical exponent which characterizes anisotropic scaling 
$x^i \to bx^{i}$ and $t \to b^z t$ at the fixed point.
The system does not possess the relativistic invariance for $z \ne 1$ but spatial rotational invariance
and translational invariance.
The Lorentz invariance is expected to emerge after the transition from $z \ne 1$ to $z = 1$.

Let us first consider a simple model on $\bm{R}^3 \times \bm{R}$ with a complex scalar field $\phi$ and two kinds of spinor fields $\varphi$ 
and $\eta$ 
described by the action:
\begin{eqnarray}
&~& S_0 = \int dt d^3{\bm{x}} 
\left[ \left|\frac{\partial \phi}{\partial t}\right|^2 - \frac{1}{\kappa^2}\left|\frac{\partial^2 \phi}{\partial x^i \partial x^i}\right|^2 \right.
\nonumber \\
&~& ~~~~~~~ \left. + \varphi^{\dagger} i \frac{\partial}{\partial t} \varphi + \eta^{\dagger} i \frac{\partial}{\partial t} \eta
+ \frac{1}{\xi^2} \varphi^{\dagger} \frac{\partial^2}{\partial x^i \partial x^i} \eta 
+ \frac{1}{\xi^2} \eta^{\dagger} \frac{\partial^2}{\partial x^i \partial x^i} \varphi \right]~.
\label{S0}
\end{eqnarray}
Both $\varphi$ and $\eta$ are  2-component spinors defined on $\bm{R}^3$ and
they transform as 
\begin{eqnarray}
&~& \varphi(x) \to \varphi' \to \varphi'(x') = S(O) \varphi(x)~, ~~ \eta(x) \to \eta' \to \eta'(x') = S(O) \eta(x)~,
\label{Rotation}\\
&~& S(O) \equiv e^{-\frac{i}{4}\omega_{ij}\sigma^{ij}}~,~~ \sigma^{ij} \equiv \frac{i}{2}\left(\sigma^i \sigma^j - \sigma^j \sigma^i \right)~,
\label{S(O)}
\end{eqnarray}
under the $3$-dimensional rotation $x^i \to x'^{i} = O^i_j x^j$.
Here, $\sigma^i$s are Pauli matrices, $\omega_{ij}$ are parameters related to rotation angles $\theta^i$
such as $\omega_{ij} = - \varepsilon_{ijk} \theta^k$
and $O^i_j$ is the $3 \times 3$ orthogonal matrix given by $\displaystyle{O^i_j = (e^{\omega})^i_j}$.
The $S(O)$ satisfies the following relations:
\begin{eqnarray}
S^{\dagger}(O) \sigma^i S(O) = O^i_j \sigma^j~, ~~ S^{\dagger}(O) S(O) = I~,
\label{S-rel}
\end{eqnarray}
where $I$ is the $2 \times 2$ unit matrix.
The terms such as $\varphi^{\dagger} \partial_i \partial_i \varphi$ and $\eta^{\dagger} \partial_i \partial_i \eta$
are also invariant under the spatial rotation, but we assume that they are absent.
Then (\ref{S0}) is rewritten as
\begin{eqnarray}
&~& S_0 = \int dt d^3{\bm{x}} 
\left[ \left|\frac{\partial \phi}{\partial t}\right|^2 - \frac{1}{\kappa^2}\left|\frac{\partial^2 \phi}{\partial x_i \partial x_i}\right|^2 
+ \bar{\psi} i \gamma^0 \frac{\partial}{\partial t} \psi + \frac{1}{\xi^2} \bar{\psi} \frac{\partial^2}{\partial x_i \partial x_i} \psi \right]~,
\label{S0-psi}
\end{eqnarray}
where we define a 4-component spinor field as $\psi \equiv (\varphi, \eta)^t$, $\bar{\psi} \equiv \psi^{\dagger} \gamma^0$ and 
$\gamma^0$ corresponds to a time component of the gamma matrices given by
\begin{eqnarray}
\gamma^0 = 
\left(
\begin{array}{cc}
0 & I \\
I & 0 
\end{array}
\right)~.
\label{gamma0}
\end{eqnarray}
The engineering dimensions of fields ($\phi$, $\psi$) and couplings ($\kappa^2$, $\xi^2$) are given by
\begin{eqnarray}
[\phi] = \frac{3-z}{2}~,~~ [\psi]=\frac{3}{2}~,~~[\kappa^2]=4-2z~,~~[\xi^2]=2-z~.
\label{dim}
\end{eqnarray}
The system has a free-field fixed point with $z=2$. 

Next let us adopt the gauge principle.
The gauge fields ($A_t$, $A_i$) are introduced and the derivatives 
($\partial_t \equiv \partial/\partial t$, $\partial_i \equiv \partial/\partial x^i$) are
replaced into the covariant ones ($D_t$, $D_i$) such as
\begin{eqnarray}
\partial_t \Rightarrow D_t \equiv \partial_t + i g A_t~,~~ 
\partial_i \Rightarrow D_i \equiv \partial_i + i g A_i~,
\label{D}
\end{eqnarray}
where $g$ is a gauge coupling.
The engineering dimensions of $A_t$ and $A_i$ are
\begin{eqnarray}
[A_t] = z-[g]~, ~~ [A_i] =1-[g]~.
\label{dimA}
\end{eqnarray}
Gauge field strengths ($F_{ti}$, $F_{ij}$) are constructed from the commutators of covariant derivatives as
\begin{eqnarray}
[D_t, D_i] = ig F_{ti}~,~~ [D_i, D_j] = ig F_{ij}~.
\label{DD}
\end{eqnarray}
If the gauge fields are dynamical, the kinetic term of gauge fields should be added.
In this way, we obtain the following gauge invariant action:
\begin{eqnarray}
&~& S = \int dt d^3{\bm{x}} \left[{\rm tr}(F_{ti})^2 -\frac{1}{2\lambda^2}{\rm tr}(D_kF_{ij})^2 \right.
\nonumber \\
&~& ~~~~~~~~~ \left. 
+ \left|D_t \phi\right|^2 - \frac{1}{\kappa^2}\left|D_iD_i \phi\right|^2 
+ \bar{\psi} i \gamma^0 D_t \psi + \frac{1}{\xi^2} \bar{\psi} D_i D_i \psi \right]~,
\label{S}
\end{eqnarray}
where tr represents the trace for the gauge generators.
If the differential operator $\partial_i \partial_i$ for $\varphi$ and $\eta$ in (\ref{S0}) is 
regarded as $(\sigma^i \partial_i)^2$, the last term in (\ref{S}) is replaced by $-\xi^{-2} \bar{\psi} (\gamma^i D_i)^2 \psi$. 
Here, $\gamma^i$s correspond to space components of gamma matrices given by
\begin{eqnarray}
\gamma^i =
\left(
\begin{array}{cc}
0 & -\sigma^i \\
\sigma^i & 0 
\end{array}
\right)~
\label{gammai}
\end{eqnarray}
and satisfy the relation $\gamma^i\gamma^j + \gamma^j \gamma^i = -2\delta^{ij}$.
We add ${\rm tr}(D_kF_{ij})^2$ as the $\lq$potential term' for $A_i$ to the action $S$
with an extended model with a higher spatial derivative in mind.
The $\lq$potential term' means terms other than the kinetic term including time derivatives.
If we use the detailed balance condition, the term proportional to ${\rm tr}(D_iF_{ik}D_jF_{jk})$ is introduced.
Here, the detailed balance condition means that the action ($S_{D+1}$)  in $D+1$ dimensions
is constructed from the action ($S_D$) in $D$ dimensions through a particular procedure
using functional derivatives.
For example, in a model with a real scalar field $\Phi$, the action is constructed as
\begin{eqnarray}
&~& S_{D+1} = \int dt d^D{\bm{x}} \left[ \frac{1}{2} \left(\frac{\partial \Phi}{\partial t} 
+ \frac{1}{2\kappa}\frac{\delta S_D}{\delta \Phi}\right)^2 \right]~,
\label{S_D+1}
\end{eqnarray}
up to total derivative term.
The dimension of $[{\rm tr}(F_{ti})^2]$ is given by $3+z$.
Then the dimensions of ($A_t$, $A_i$) and ($g$, $\lambda^2$) are
\begin{eqnarray}
[A_t] = \frac{1+z}{2}~, ~~ [A_i] =\frac{3-z}{2}~,~~[g]=\frac{z-1}{2}~,~~[\lambda^2]=4-2z~.
\label{dimAglambda}
\end{eqnarray}

The effective field theory, in general, contains operators whose dimensions equal to or less than those in $S$,
which are not forbidden by symmetry.
We add the following relevant terms directly to $S$:
\begin{eqnarray}
\hspace{-0.5cm}\Delta S 
= \int dt d^3{\bm{x}} \left[  - \frac{{\tilde{c}_A}^{2z-2}}{2}{\rm tr} (F_{ij})^2 
 - {\tilde{c}_{\phi}}^{2z-2} |D_i \phi|^2  + {\tilde{c}_{\psi}}^{z-1} \bar{\psi} i \gamma^i D_i \psi + \cdots \right]~,
\label{DeltaS}
\end{eqnarray}
where $\tilde{c}_A$, $\tilde{c}_{\phi}$ and $\tilde{c}_{\psi}$ are parameters with dimension one
and the ellipsis stands for terms related to other operators such as 
Yukawa interactions and self-interactions of scalar fields.
The third term in (\ref{DeltaS}) is written in terms of 2-component spinors $\varphi$ and $\eta$ as
\begin{eqnarray}
{\tilde{c}_{\psi}}^{z-1} \bar{\psi} i \gamma^i D_i \psi = {\tilde{c}_{\psi}}^{z-1} \left(\varphi^{\dagger}i\sigma^i D_i \varphi
 -\eta^{\dagger}i\sigma^i D_i \eta \right)~.
\label{DeltaS-psi}
\end{eqnarray}

It is straightforward to construct a model with a complex scalar field $\phi$, a $2^{[(D+1)/2]}$-component spinor field $\psi$ and 
gauge fields $(A_t, A_i)$ on $\bm{R}^D \times \bm{R}$. 
The superficial degree of divergence ($D_s$) is defined by
\begin{eqnarray}
D_s \equiv D+z -\frac{D-z}{2} B - \left(\frac{D+z}{2}-1\right) B_t - \frac{D}{2} F - \sum_k [g^k] n_k - \cdots~,
\label{Ds}
\end{eqnarray}
where $B$, $B_t$, $F$ and $n_k$ are numbers of external lines for $\phi$ and $A_i$, 
numbers of external lines for $A_t$, external lines for $\psi$ and vertices including the $k$-th power of $g$.
The ellipsis represents contributions from other couplings.

In the case with $z=2$ and $D=3$,
the dimensions of fields and couplings become
\begin{eqnarray}
&~& [\phi] = \frac{1}{2}~,~~ [\psi]=\frac{3}{2}~,~~[A_t] = \frac{3}{2}~,~~ [A_i] =\frac{1}{2}~,
\label{dimfields}\\
&~& [g]=\frac{1}{2}~,~~[\kappa^2]=[\xi^2]=[\lambda^2]=0~,
\label{dimcoupling}
\end{eqnarray}
and $D_s$ is given by
\begin{eqnarray}
D_s = 5 -\frac{1}{2} B - \frac{3}{2} B_t - \frac{3}{2} F - \frac{1}{2}\sum_k k n_k~.
\label{Dsz=2}
\end{eqnarray}
The theory is superrenormalizable by power counting and
infinities appear for a finite number of diagrams if exist.
In Appendix A, we study radiative corrections at one-loop level using an abelian gauge theory
and find that radiative corrections for gauge coupling do not contain any infinities owing to the gauge invariance.
We expect that the same property holds on the Lifshitz type extension of the MSSM and SUSY GUT.

\subsection{A chiral model}

On the construction of GUT or an intermediate theory, a simple group or the SM gauge group 
should be chosen as a gauge group with suitable particle contents (including chiral fermions).
Here, we do not specify the gauge group and particle contents for simplicity.
We discuss a model with $z=3$ on the space-time $\bm{R}^3 \times \bm{R}$, including chiral fermions with a higher spatial derivative term.
A candidate is give by the action:
\begin{eqnarray}
&~& S_{z=3} = S + \Delta S~,
\nonumber \\
&~& S = \int dt d^3{\bm{x}} \left[{\rm tr}(F_{ti})^2 -\frac{1}{2\lambda^2}{\rm tr}(D_kD_lF_{ij})^2 
+ \sum_{f} \left({\varphi}_f^{\dagger} i D_t \varphi_f - \frac{1}{\xi^2} \varphi_f^{\dagger} {\bm{D}}^2 i \sigma^i D_i \varphi_f \right) \right.
\nonumber \\
&~& ~~~~~~~~~~~~~~~~~~~~~ \left.
+ \sum_{h} \left(\left|D_t \phi_h\right|^2 - \frac{1}{\kappa^2}\left|{\bm{D}}^2 D_i \phi_h\right|^2\right) \right]~,
\nonumber \\
&~&  \Delta S =  \int dt d^3{\bm{x}} \left[ - \frac{{\tilde{c}_A}^{4}}{2} {\rm tr} (F_{ij})^2 
  + \sum_{f} {\tilde{c}_{f}}^{2} \varphi_f^{\dagger} i \sigma^i D_i \varphi_f - \sum_{h} {\tilde{c}_{h}}^{4} |D_i \phi_h|^2 + \cdots \right]~,
\label{Sz=3}
\end{eqnarray}
where $\varphi_f$ are 2-component spinor fields, $\phi_h$ are scalar fields, ${\bm{D}}^2 = D_i D_i$
and the ellipsis stands for terms related to other operators such as 
Yukawa interactions and self-interactions of scalar fields.
We can, in general, add various kinds of gauge invariant operators whose dimension equals to or less than 6.
Here, we take simple ones in $S_{z=3}$.
For example, we assume that operators such as $\varphi_f^{\dagger} \varphi_f$ and $\varphi_f^{\dagger} {\bm{D}}^2 \varphi_f$
are absent and do not appear through radiative corrections.
If $\varphi_f^{\dagger} \varphi_f$ exists, the Lorentz invariance does not appear in the IR region.
We expect that the above type of action without specific operators is derived from a more fundamental theory.

The engineering dimensions of fields and parameters are given by
\begin{eqnarray}
&~& [A_t] = 2~,~~ [A_i] =0~,~~[\varphi_f]=\frac{3}{2}~,~~[\phi_h]=0~, 
\label{dim-fields}\\
&~& [g]=1~,~~[\lambda^2]=[\xi^2]=[\kappa^2]=0~,~~[\tilde{c}_A] = [\tilde{c}_{f}] = [\tilde{c}_h]=1~
\label{dim-parameters}
\end{eqnarray}
and $D_s$ is given by
\begin{eqnarray}
D_s = 6 - 2 B_t - \frac{3}{2} F - \sum_k k n_k~.
\label{Dsz=2}
\end{eqnarray}
This theory is also superrenormalizable by power counting.
It is important to study the ultraviolet (UV) behavior of our model more carefully.

{}From a naive dimensional analysis, $\Delta S$ dominates against $\lq$potential terms' in $S$ 
at the energy scale below $\tilde{c}_A$, $\tilde{c}_{f}$ and $\tilde{c}_{h}$,
and we expect that the relativistic invariance appears there and
the transition occurs from the theory with $z = 3$ to that with $z = 1$.
In order to become relativistic below $M_X$,
a finetuning among parameters is required such that $\tilde{c}_A = \tilde{c}_{f} = \tilde{c}_{h} \equiv M_{\ell}$
to a high accuracy and the order of $M_{\ell}$ is (one or two digit) bigger than that of $M_X$.
We will discuss constraints for these parameters from experiments in the next subsection.
This relation could be realized as renormalization group (RG) invariants.\footnote{
There has been a proposal that the Lorentz invariance appears at an attractive IR fixed point.\cite{NN}
In Ref.~\citen{IRS}, the RG evolution of $\tilde{c}_{h}$ has been studied 
for the Lifshitz type scalar field theory with $z=2$ and $D=4$ ($D=10$)
and it has been pointed out that a severe fine-tuning in the UV region seems to be inevitable.
This result can give a strong constraint on theories with extra dimensions.
}
Then $M_{\ell}^2 t$ is regarded as an ordinary time variable $x_0$ in the relativity.
Hence we expect that each term in $\Delta S$ comes from a common origin, 
$\tilde{c}_A$, $\tilde{c}_{f}$ and $\tilde{c}_{h}$ 
take a common value at the beginning and they do not receive radiative corrections.
If the finetuning relation holds on for all particles and 
the theory has suitable particle contents (gauge bosons, chiral fermions, Higgs bosons and their superpartners)
with suitable gauge quantum numbers,
the MSSM can be derived as the theory below $M_X$ and
the running of gauge couplings in the IR region is understood.

\subsection{Constraints from experiments}

In the Lifshitz type extension of the MSSM with $z >1$,
the dispersion relations for free fields are given by
\begin{eqnarray}
E^2 = {\bm{p}}^2 c_k^2  + {m_{k}}^2 c_k^4  + \frac{\zeta_k}{M_{\ell}^{2z-2}} {\bm{p}}^{2z} c_k^{4-2z}~,
\label{dis-rel}
\end{eqnarray}
where ${\bm{p}}^{2z}=({\bm{p}}^2)^z$, $c_k= c(\tilde{c}_k/M_{\ell})^{z-1}$ ($c$ is a speed of light),
$\zeta_k$ are dimensionless parameters and
every particle (except gauge bosons related to unbroken gauge symmetries) acquires
the mass $m_k$ after the SUSY and/or electroweak symmetry breaking.
Here, $c_k$ is an $\lq$own velocity' (a maximal attainable velocity if $\zeta_k > 0$) for each particle labeled by $k$.

Let us discuss constraints for parameters from experimental data.
The Lorentz violating dispersion relations such as (\ref{dis-rel}) are tested by the photon emission
from a high-energy particle.\cite{Lvio-test}
The process such as $C \to C + \gamma$ is forbidden (at tree level) by the energy-momentum conservation
in the case with the Lorentz invariant dispersion relations such as $E^2 = {\bm{p}}^2 c^2  + {m_{C}}^2 c^4$
for a massive charged particle $C$ and $E^2 = {\bm{p}}^2 c^2$ for photon $\gamma$.
It, however, can occur in the presence of Lorentz symmetry breaking terms.
We consider the case with the following dispersion relations:
\begin{eqnarray}
&~& E^2 = {\bm{p}}^2 c_C^2 + {m_{C}}^2 c_C^4 + \frac{\zeta_C}{M_{\ell}^{n-2}} {\bm{p}}^n c_C^{4-n} ~,
\label{dis-rel-C}\\
&~& E^2 = {\bm{p}}^2 c_{\gamma}^2  + \frac{\zeta_{\gamma}}{M_{\ell}^{n-2}} {\bm{p}}^n c_{\gamma}^{4-n} ~,
\label{dis-rel-gamma}
\end{eqnarray}
where $n = 2z$.
Using the kinematics, we derive the following relation:
\begin{eqnarray}
&~& \frac{\zeta_{\gamma}}{M_{\ell}^{n-2}} {\bm{p}}_{\gamma}^n c_{\gamma}^{4-n} 
= \left({\bm{p}}^2 + {\bm{p}}'^2\right) \left(c_C^2 - c_{\gamma}^2\right)
\nonumber \\
&~& ~~~~~~~ + 2\left( {m_{C}}^2 c_C^4 + {\bm{p}}\cdot{\bm{p}}' c_{\gamma}^2 - EE' \right)
+ \frac{\zeta_C}{M_{\ell}^{n-2}} ({\bm{p}}^n +  {\bm{p}}'^n) c_C^{4-n} ~,
\label{kinematics}
\end{eqnarray}
where ${\bm{p}}_{\gamma}$ is the momentum of photon, ${\bm{p}}$ (${\bm{p}}'$) and $E$ ($E'$) are 
the momentum and energy of incoming (outgoing) particle $C$.
$E$ and $E'$ satisfy the relation (\ref{dis-rel-C}) and the same type one whose ${\bm{p}}$ is replaced by ${\bm{p}}'$, respectively.
The stability of $C$ at some high-momentum ${\bm{p}}_{\tiny{\mbox{h}}}$ means that the relation (\ref{kinematics})
does not hold on at and below ${\bm{p}}_{\tiny{\mbox{h}}}$ and it leads to the following constraints:
\begin{eqnarray}
\frac{|c_C^2 - c_{\gamma}^2|}{c_C^2} \lesssim \frac{m_C^2 c_C^2}{{\bm{p}}_{\tiny{\mbox{h}}}^2}~,~~ 
\zeta_k \lesssim \frac{m_C^2 M_{\ell}^{n-2}}{{\bm{p}}_{\tiny{\mbox{h}}}^n}c_k^n~,
\label{C-cons}
\end{eqnarray}
where $k$ is $C$ or $\gamma$.
We use the relation $c_C^2 = c_{\gamma}^2$ up to a very tiny number expected from the first constraint to derive the second one.
The most stringent constraints come from the case that $C$ is proton ($p^+$)\footnote{
Strictly speaking, proton does not belong to the SM particles as an elementary particle
and hence $\zeta_p$ would be zero in our Lifshitz type model.}
and they are given by\footnote{
Various constraints on maximal attainable velocities for particles have been derived in Ref.~\citen{C&G}.}
\begin{eqnarray}
\frac{|c_p^2 - c_{\gamma}^2|}{c_p^2} \lesssim \frac{m_p^2 c_p^2}{{\bm{p}}_{\tiny{\mbox{h}}}^2} = O(10^{-22}) ~,~~ 
\zeta_k \lesssim \frac{m_p^2 M_{\ell}^{n-2}}{{\bm{p}}_{\tiny{\mbox{h}}}^n}c_k^n = O(10^{6n-34})~,
\label{p-cons}
\end{eqnarray}
where we use ${\bm{p}}_{\tiny{\mbox{h}}} = 10^{20}$eV from AGASA date\cite{AGASA}
and $m_p = 938$MeV.
We take $M_{\ell} = 10^{17}$GeV.
The first constraint requires that $\tilde{c}_p = \tilde{c}_{\gamma}$ to a high accuracy of order $O(10^{-22}M_{\ell})$.
{}From the second one,  we find that the magnitude of $\zeta_k$ can become $O(1)$ if $n \geq 6$.
In the case that $C$ is electron ($e^-$), constraints are given by
\begin{eqnarray}
\hspace{-0.5cm} \frac{|c_e^2 - c_{\gamma}^2|}{c_e^2} \lesssim \frac{m_e^2 c_e^2}{{\bm{p}}_{\tiny{\mbox{h}}}^2} = O(10^{-16})~,~~ 
\zeta_k \lesssim \frac{m_e^2 M_{\ell}^{n-2}}{{\bm{p}}_{\tiny{\mbox{h}}}^n}c_k^n = O(10^{12n-40}/2^{2-n})~,
\label{e-cons}
\end{eqnarray}
where we use ${\bm{p}}_{\tiny{\mbox{h}}} = 50$TeV from observations of high-momentum gamma ray 
originated from the Crab nebula\cite{Crab}
and $m_e = 0.51$MeV.
We take $M_{\ell} = 10^{17}$GeV.
The first constraint requires that $\tilde{c}_e = \tilde{c}_{\gamma}$ to a high accuracy of order $O(10^{-16}M_{\ell})$.
{}From the second one,  we find that the magnitude of $\zeta_k$ can become $O(1)$ if $n \geq 4$.

If the magnitude of Lorentz symmetry breaking term can be within the range of observation
concerning photon with an ultra high-momentum
in the future, we come across the idea that
{\it light led us to the concept of special relativity, but (ironically) it also might teach us to 
the violation of relativistic invariance and a new physics beyond the SM or the MSSM.}

\section{Conclusions and discussion}

We have proposed a new grand unification scenario for ensuring proton stability 
on the basis of the conjecture that {\it the proton stability can be realized to be compatible with the grand unification
if $M_X$ is not related to $M_U$} and the standpoint that {\it the scale $M_X$ has a physical meaning
beyond the fact that the agreement of gauge couplings occurs accidentally there}.
In our scenario, $M_U$ is supposed to be bigger than $M_X$ to suppress the prodon decay sufficiently 
and gauge couplings agree with among them from $M_U$ to $M_X$ by a specific dynamics.
The candidate is a Lifshitz type gauge theory with $z>1$, 
which possesses specific superrenormalizable interactions.
The structure of theory changes around $M_X$ after a phase transition irrelevant to the grand unification
and the different running of $g_i$ starts there.
The transition occurs from the theory with $z>1$ to that with $z=1$,
and the ordinary renormalizable MSSM is expected to hold on with the emergence of Poincare invariance and SUSY
below $M_X$.

There are open questions for our scenario and/or model.
The most serious problem is a finetuning problem related to the speed of light.
The relativistic invariance requires a finetuning among parameters such as
$\tilde{c}_A = \tilde{c}_{f} = \tilde{c}_{h} \equiv M_{\ell}$ to a high accuracy
for all particles that survive in our low-energy world.
This comes from the standpoint that the Lorentz invariance emerges as an accidental symmetry.
The relation could be realized as RG invariants.
Then the RG invariance (or no running of parameters) would become one of conditions
to select a realistic model.
We need to study radiative corrections or RG behavior 
in the space of all the couplings in our model, bearing the finetunig problem in mind. 
Furthermore $M_{\ell}$ might be a fundamental constant in a fundamental theory, e.g. the string scale in string theory.
In fact, it is known that the string scale is about 20 times bigger than $M_X$ in the heterotic string theory.
Or we might need to reconsider the relation between the kinematics of each particle and the structure of space-time.
Other notorious problem is the triplet-doublet Higgs mass splitting problem.
There is a possbility that the mass splitting can be realized
on the orbifold breaking\cite{orb} if a higher-dimensional gauge theory holds on above $M_U$.
It is interesting to study this problem in our framework.
It is also important to explore phenomenological implications on
the violation of Lorentz invariance at a high-energy scale,
considering phenomenological aspect of Ho\v{r}ava-Lifshitz gravity.\footnote{
Recently, various aspects of Ho\v{r}ava-Lifshitz gravity have been  intensively investigated
from its renormalizability to cosmological implications.\cite{HoravaGr}}

Even if our present model do not work for any reason, study of elementary particle physics using
Lifshitz type quantum field theories would be attractive
and it is worth exploring a high-energy theory on the basis of our proposal.
In any event, we would like to believe that nature use the idea of grand unification
in order to well-organize physical laws.\footnote{
Recently, unconventional idea that grand unification is realized in unphysical world has been proposed.\cite{TGU}}$^{,}$
\footnote{
On the other hand, there is an interesting proposal that the merging of gauge couplings can come from not the grand unification 
but the property of an underlying fermionic vacuum and there is no danger of proton decay.\cite{K&V}
}

\section*{Acknowledgements}
This work was supported in part by scientific grants from the Ministry of Education, Culture,
Sports, Science and Technology under Grant Nos.~18204024 and 18540259.

\appendix

\section{Radiative Corrections in Lifshitz Type Abelian Gauge Theory}

We study radiative corrections at one-loop level using a Lifshitz type abelian gauge theory
with $z=2$ on the space-time $\bm{R}^3 \times \bm{R}$.
Our players are a $U(1)$ gauge field ($A_t$, $A_i$) and a fermion $\psi$,
which is a 4-component spinor (defined on $\bm{R}^3$) and has a unit $U(1)$ charge $e$.
The action is given by
\begin{eqnarray}
&~& S = \int dt d^3{\bm{x}} \left[\frac{1}{2}(F_{ti})^2 -\frac{1}{4}(\partial_kF_{ij})^2
 + \bar{\psi} i \gamma^0 D_t \psi + \bar{\psi} D_i D_i \psi \right.
\nonumber \\
&~& ~~~~~~~~~~~~~~~~~~~ \left. - \frac{{\tilde{c}_A}^{2}}{4} (F_{ij})^2 
 + \tilde{c}_{\psi} \bar{\psi} i \gamma^i D_i \psi - {\tilde{m}}^2 \bar{\psi} \psi \right]~,
\label{S-QED}
\end{eqnarray}
where $F_{ti} = \partial_t A_i - \partial_i A_t$, $F_{ij} = \partial_i A_j - \partial_j A_i$,
$D_t \equiv \partial_t + i e A_t$ and $D_i \equiv \partial_i + i e A_i$.
The engineering dimensions of fields and parameters are given by
\begin{eqnarray}
[A_t] = \frac{3}{2}~,~~ [A_i] =\frac{1}{2}~,~~[\psi]=\frac{3}{2}~,~~
[e]=\frac{1}{2}~,~~[\tilde{c}_A] = [\tilde{c}_{\psi}] = [\tilde{m}]=1~.
\label{dimQED}
\end{eqnarray}
The Feynman rules are given by
\begin{eqnarray}
&~& \frac{i}{\gamma^0 E_p - {\bm{p}}^2 - \tilde{c}_{\psi} {\bm{\gamma}}\cdot{\bm{p}}- \tilde{m}^2 + i\varepsilon}
~~~~ {\mbox{for propagator of $\psi$}}~,\\
&~& \frac{-i({\bm{p}}^2+\tilde{c}_A^2)}{E_p^2 - {\bm{p}}^4 - \tilde{c}_A^2 {\bm{p}}^2 + i\varepsilon}~~~~~~~~~~~~~~ {\mbox{for propagator of $A^t$}}~,\\
&~& \frac{i}{E_p^2 - {\bm{p}}^4 - \tilde{c}_A^2 {\bm{p}}^2 + i\varepsilon}~~~~~~~~~~~~~~ {\mbox{for propagator of $A^i$}}~,\\
&~& -ie\gamma^0 ~~~~~~~~~~~~~~~~~~~~~~~~~~~~~~~~~~~ {\mbox{for vertex among $\bar{\psi}$, $\psi$ and $A^t$}}~,\\
&~& -ie(p_i + p'_i + \tilde{c}_{\psi} \gamma_i)~~~~~~~~~~~~~~~~ {\mbox{for vertex among $\bar{\psi}$, $\psi$ and $A^i$}}~,\\
&~& - 2ie^2 \delta_{ij}~~~~~~~~~~~~~~~~~~~~~~~~~~~~~~~ {\mbox{for vertex among $\bar{\psi}$, $\psi$, $A^i$ and $A^j$}}~,
\label{Frule}
\end{eqnarray}
where $\displaystyle{{\bm{p}}^4 \equiv ({\bm{p}}^2)^2}$ and we take $\partial_t A_t + (\nabla^2 - \tilde{c}_A^2) \partial_i A_i = 0$
as the gauge fixing condition to derive the propagators of $A_t$ and $A_i$.

Now let us study radiative corrections at one-loop level.
Here, we do not consider the terms $\displaystyle{\frac{{\tilde{c}_A}^{2}}{4} (F_{ij})^2}$ and $\tilde{c}_{\psi} \bar{\psi} i \gamma^i D_i \psi$
for simplicity.\footnote{
The qualitative features are almost same by the introduction of those terms.
We will report results of analysis including those terms elsewhere.}
The vacuum polarizations at one-loop level are given by
\begin{eqnarray}
&~& i\Pi_{tt}(q)
= -e^2 \int_{-\infty}^{\infty} \frac{dE_p d^3{\bm{p}}}{(2\pi)^4}\mbox{Tr} \left(\gamma^0 \frac{1}{\gamma^0 E_p - {\bm{p}}^2 - \tilde{m}^2}\right.
\nonumber \\
&~& ~~~~~~~~~~~~~~~~~~~~~~~~~~~~~~~~~~~~~~~~  \left. \cdot \gamma^0 \frac{1}{\gamma^0 (E_p -E_q) - ({\bm{p}}-{\bm{q}})^2- \tilde{m}^2}\right)~,
\label{Pitt}\\
&~& i\Pi_{it}(q) = - e^2 \int_{-\infty}^{\infty} \frac{dE_p d^3{\bm{p}}}{(2\pi)^4} \mbox{Tr} \left(\frac{2p_i -q_i}{\gamma^0 E_p - {\bm{p}}^2- \tilde{m}^2}\right.
\nonumber \\
&~& ~~~~~~~~~~~~~~~~~~~~~~~~~~~~~~~~~~~~~~~~  \left. \cdot 
\gamma^0 \frac{1}{\gamma^0 (E_p -E_q) - ({\bm{p}}-{\bm{q}})^2- \tilde{m}^2}\right)~,
\label{Piit}\\
&~& i\Pi_{ij}(q) = -e^2 \int_{-\infty}^{\infty} \frac{dE_p d^3{\bm{p}}}{(2\pi)^4} \left[\mbox{Tr} \left(\frac{2p_i -q_i}{\gamma^0 E_p - {\bm{p}}^2- \tilde{m}^2}
\right.\right.
\nonumber \\
&~& ~~~~~~~~~~~~~~~~~~~~~~~~~~~~~~~~~~~~~~~~  \left.
\cdot \frac{2p_j-q_j}{\gamma^0 (E_p -E_q) - ({\bm{p}}-{\bm{q}})^2- \tilde{m}^2}\right) \nonumber \\
&~& ~~~~~~~~~~~~~~~~~~~~~~~~~~~~~~~~~~~ + \left. \mbox{Tr} \left(\frac{2\delta_{ij}}{\gamma^0 E_p - {\bm{p}}^2- \tilde{m}^2}\right) \right]~,
\label{Piij}
\end{eqnarray}
where $q=(E_q, {\bm{q}})$ ($E_q$ and ${\bm{q}}$ are energy and momenta for the external gauge boson, respectively)
and Tr represents the trace for the spinor indices.
$\Pi_{tt}(q)$, $\Pi_{it}(q)$ and $\Pi_{ij}(q)$ seem contain linear, quadratic and tertiary divergent pieces,
but they can be removed from the condition of gauge invariance.
Physics must not be changed under the gauge transformations
$A_t \to A_t + \partial_t \Lambda$ and $A_i \to A_i + \partial_i \Lambda$
($\varepsilon_t \to \varepsilon_t + {\lambda} E_q$ and $\varepsilon_i \to \varepsilon_i + {\lambda} q_i$
in terms of polarization vectors $\varepsilon_t$ and $\varepsilon_i$).
This requirement leads to 
the relations $E_q \Pi_{tt} - \sum_{i} q_i \Pi_{it} = 0$ and $E_q \Pi_{it} - \sum_{j} q_j \Pi_{ij} = 0$.
Using the above expressions (\ref{Pitt}) -- (\ref{Piij}), we derive the following relations:
\begin{eqnarray}
&~& E_q \Pi_{tt} - \sum_{i} q_i \Pi_{it} = i e^2 \int_{-\infty}^{\infty} \frac{dE_p d^3{\bm{p}}}{(2\pi)^4}
\mbox{Tr} \left(\gamma^0\frac{1}{\gamma^0 (E_p -E_q) - ({\bm{p}}-{\bm{q}})^2 - \tilde{m}^2} \right.
\nonumber \\
&~& ~~~~~~~~~~~~~~~~~~~~~~~~~~~~~~~~~~~~~~~~~~~~~~~~~~~~~~~~~~~ \left. - \gamma^0\frac{1}{\gamma^0 E_p - {\bm{p}}^2- \tilde{m}^2}\right)~,
\label{gaugeinv1}\\
&~& E_q \Pi_{it} - \sum_{j} q_j \Pi_{ij} = i e^2 \int_{-\infty}^{\infty} \frac{dE_p d^3{\bm{p}}}{(2\pi)^4}
\mbox{Tr} \left(\frac{2p_i - q_i}{\gamma^0 (E_p -E_q) - ({\bm{p}}-{\bm{q}})^2- \tilde{m}^2} \right.
\nonumber \\
&~& ~~~~~~~~~~~~~~~~~~~~~~~~~~~~~~~~~~~~~~~~~~~~~~~~~~~~~~~~~~~ \left. - \frac{2p_i+q_i}{\gamma^0 E_p - {\bm{p}}^2- \tilde{m}^2}\right)~.
\label{gaugeinv2}
\end{eqnarray}
If the integrals were finite, we could make them zero after the change of integration variables
in the first terms.
Actually we can regularize $\Pi_{tt}(q)$, $\Pi_{it}(q)$ and $\Pi_{ij}(q)$ keeping the gauge invariance
and they have the following forms:
\begin{eqnarray}
&~& \Pi_{tt}(q)= {\bm{q}}^2 \Pi(q)~,~~\Pi_{it}(q)= E_q q_i \Pi(q)~,\nonumber \\
&~& \Pi_{ij}(q)= E_q^2 \delta_{ij} \Pi(q) + \left({\bm{q}}^2 \delta_{ij} - q_i q_j\right) \tilde{\Pi}(q)~,
\label{Pi}
\end{eqnarray}
where $\Pi(q)$ is a finite quantity.
For example, $i\Pi(0)$ is calculated as
\begin{eqnarray}
&~& i \Pi(0) = - e^2 \int_{-\infty}^{\infty} \frac{dE_p d^3{\bm{p}}}{(2\pi)^4}
\mbox{Tr} \left(\frac{1}{(\gamma^0 E_p - ({\bm{p}}^2 + \tilde{m}^2))^3} \right)
\nonumber \\
&~& ~~~~~~~ = - e^2 \int_{-\infty}^{\infty} \frac{dE_p d^3{\bm{p}}}{(2\pi)^4}
\mbox{Tr} \left(\frac{(\gamma^0 E_p + ({\bm{p}}^2 + \tilde{m}^2))^3}{(E_p^2 - ({\bm{p}}^2 + \tilde{m}^2)^2)^3} \right)
\nonumber \\
&~& ~~~~~~~ = - 4i e^2 \int_{-\infty}^{\infty} \frac{dE d^3{\bm{p}}}{(2\pi)^4}
\frac{-3 E^2 ({\bm{p}}^2 + \tilde{m}^2)+({\bm{p}}^2 + \tilde{m}^2)^3}{(E^2 + ({\bm{p}}^2 + \tilde{m}^2)^2)^3} 
\nonumber \\
&~& ~~~~~~~ = - 4i e^2 \int_{-\infty}^{\infty} \frac{d^3{\bm{p}}}{(2\pi)^4}
\left(\frac{-3}{8}\frac{\pi}{{\bm{p}}^2 + \tilde{m}^2} + \frac{3}{8}\frac{\pi}{{\bm{p}}^2 + \tilde{m}^2}\right)
= 0~,
\label{Pi0}
\end{eqnarray}
where the Wick rotation is performed ($E_p \to iE$).

In the same way, we can calculate the fermion self-energy and the vertex functions.
The expression for the fermion self-energy is given at one-loop level,
\begin{eqnarray}
&~& -i \Sigma(q)
= e^2 \int_{-\infty}^{\infty} \frac{dE_p d^3{\bm{p}}}{(2\pi)^4}
\frac{(2{\bm{q}}-{\bm{p}})^2}{E_p^2 - {\bm{p}}^4} \frac{1}{\gamma^0 (E_q -E_p) - ({\bm{q}}-{\bm{p}})^2- \tilde{m}^2}
\nonumber \\
&~& ~~~~~~~~~~~~~~~~~~~~~~~ + 6 e^2 \int_{-\infty}^{\infty} \frac{dE_p d^3{\bm{p}}}{(2\pi)^4} \frac{1}{E_p^2 - {\bm{p}}^4}
\nonumber \\
&~& ~~~~~~~~~~~~~~ + e^2 \int_{-\infty}^{\infty} \frac{dE_p d^3{\bm{p}}}{(2\pi)^4}
\frac{-{\bm{p}}^2}{E_p^2 - {\bm{p}}^4} \frac{1}{\gamma^0 (E_q -E_p) - ({\bm{q}}-{\bm{p}})^2- \tilde{m}^2}~,
\label{Sigma}
\end{eqnarray}
where $E_q$ and $q_i$ are energy and momemta for the external fermion.
The expressions for vertex functions among $\bar{\psi}$, $\psi$ and $A^t$ ($A^i$) are given at one-loop level,
\begin{eqnarray}
&~&-i \Lambda_t(q,q')
= e^2 \int_{-\infty}^{\infty} \frac{dE_p d^3{\bm{p}}}{(2\pi)^4}
\gamma^0 \frac{1}{E_p^2 - {\bm{p}}^4} \frac{(2{\bm{q}}-{\bm{p}}) \cdot (2{\bm{q}'}-{\bm{p}})}{\gamma^0 (E_q -E_p) - ({\bm{q}}-{\bm{p}})^2- \tilde{m}^2}
\nonumber \\
&~& ~~~~~~~~~~~~~~~~~~~ \cdot \frac{1}{\gamma^0 (E_{q'} -E_{p}) - ({\bm{q'}}-{\bm{p}})^2- \tilde{m}^2}
\nonumber \\
&~& ~~~~~~~~~~~ + e^2 \int_{-\infty}^{\infty} \frac{dE_p d^3{\bm{p}}}{(2\pi)^4}
\gamma^0 \frac{-{\bm{p}}^2}{E_p^2 - {\bm{p}}^4} \frac{1}{\gamma^0 (E_q -E_p) - ({\bm{q}}-{\bm{p}})^2- \tilde{m}^2}
\nonumber \\
&~& ~~~~~~~~~~~~~~~~~~~ \cdot \frac{1}{\gamma^0 (E_{q'} -E_{p}) - ({\bm{q}}-{\bm{p}})^2- \tilde{m}^2}~,
\label{Lambda-t}\\
&~& -i \Lambda_i(q,q')
= e^2 \int_{-\infty}^{\infty} \frac{dE_p d^3{\bm{p}}}{(2\pi)^4}
\frac{1}{E_p^2 - {\bm{p}}^4} \frac{(2{\bm{q}}-{\bm{p}}) \cdot (2{\bm{q}'}-{\bm{p}})}{\gamma^0 (E_q -E_p) - ({\bm{q}}-{\bm{p}})^2- \tilde{m}^2}
\nonumber \\
&~& ~~~~~~~~~~~~~~~~~~~ \cdot \frac{(q+q'-2p)_i}{\gamma^0 (E_{q'} -E_{p}) - ({\bm{q'}}-{\bm{p}})^2- \tilde{m}^2}
\nonumber \\
&~& ~~~~~~~~ + 2 e^2 \int_{-\infty}^{\infty} \frac{dE_p d^3{\bm{p}}}{(2\pi)^4}
\frac{(2q-p)_i}{E_p^2 - {\bm{p}}^4} \frac{1}{\gamma^0 (E_{q} -E_p) - ({\bm{q}}-{\bm{p}})^2- \tilde{m}^2}
\nonumber \\
&~& ~~~~~~~~ + 2 e^2 \int_{-\infty}^{\infty} \frac{dE_p d^3{\bm{p}}}{(2\pi)^4}
\frac{(2q'-p)_i}{E_p^2 - {\bm{p}}^4} \frac{1}{\gamma^0 (E_{q'} -E_p) - ({\bm{q'}}-{\bm{p}})^2- \tilde{m}^2}
\nonumber \\
&~& ~~~~~~~~ + e^2 \int_{-\infty}^{\infty} \frac{dE_p d^3{\bm{p}}}{(2\pi)^4}
\frac{-{\bm{p}}^2}{E_p^2 - {\bm{p}}^4} \frac{(q+q'-2p)_i}{\gamma^0 (E_q -E_p) - ({\bm{q}}-{\bm{p}})^2- \tilde{m}^2}
\nonumber \\
&~& ~~~~~~~~~~~~~~~~~~~~ \cdot \frac{1}{\gamma^0 (E_{q'} -E_{p}) - ({\bm{q'}}-{\bm{p}})^2- \tilde{m}^2}~,
\label{Lambda-i}
\end{eqnarray}
where $E_q$ and $q_i$ ($E'_q$ and $q'_i$) are energy and momemta for the incoming (outgoing)  fermion.
We find that the above vertex functions are UV finite.
Furthermore, there exist following identities:
\begin{eqnarray}
\Lambda_t(q,q) = - \frac{\partial}{\partial E_q} \Sigma(q)~,~~ \Lambda_i(q,q) = \frac{\partial}{\partial {q_i}} \Sigma(q)~.
\label{W-id}
\end{eqnarray}
Those are counterparts of Ward identity in QED, which guarantee that the fermion self-energy and
the three-point vertex functions do not contribute the renormalization of gauge coupling by the cancellation.
We find that $D_s < 0$ for diagrams including loops related to the vertex among $\bar{\psi}$, $\psi$, $A^i$ and $A^j$
and the proper vertex functions among them are also UV finite.

\end{document}